\documentstyle[twoside]{article}
\begin{document}
\date{October 22, 1998}

\title{Direct generation of spinning Einstein--Maxwell fields\\ from static
fields\footnote{Talk presented at the 4$^{th}$ Alexander Friedmann
International Seminar on Gravitation and Cosmology, Saint Petersburg,
Russia, June 17-25 1998}}
\author{
G\'erard Cl\'ement\thanks{E-mail: gecl@ccr.jussieu.fr.}\\
{\small Laboratoire de Gravitation et Cosmologie Relativistes,}\\
{\small Universit\'e Pierre et Marie Curie, CNRS/UPRESA 7065,}\\
{\small Tour 22-12, Bo\^{\i}te 142, 4 place Jussieu, 75252 Paris cedex 05,
France}}
\maketitle

\noindent {\bf Abstract.} I present a new method to generate rotating
solutions of the Einstein--Maxwell equations from static solutions, and
briefly discuss its general properties.

\bigskip
\bigskip
The four--dimensional stationary Einstein--Maxwell equations are
well--known to be invariant under an SU(2,1) group of transformations
\cite{nk}. Applied to asymptotically flat solutions, these
transformations map monopole solutions into monopole solutions (e.g.\ the
Schwarzschild  solution into the Reissner--Nordstr\"{o}m solution), so
that they cannot be used to transform a static monopole solution into a
rotating monopole--dipole solution, such as the Kerr solution.  The
situation becomes different in the case of stationary axisymmetric
solutions, with two commuting Killing vectors. By combining the invariance
transformations associated with a given direction in the Killing 2--plane
with rotations in this plane, the infinite--dimensional Geroch group
emerges \cite{geroch}.  These transformations allow in principle the
generation of all solutions of the stationary axisymmetric
Einstein--Maxwell problem, which is thus completely integrable.  This
generation of stationary axisymmetric solutions can be achieved in a
variety of manners, for instance by exponentiation of the infinitesimal
group action \cite{kc} or by the inverse--scattering transform method
\cite{bz}.

In this talk, I shall present a new, simple method to generate
asymptotically flat spinning Einstein--Maxwell fields from asymptotically
flat static fields by using finite combinations of SU(2,1) transformations
and global coordinate transformations mixing the two Killing vectors
\cite{kerr}. After recalling briefly the Ernst approach to the reduction
of the stationary Einstein--Maxwell problem, I shall describe this direct
rotation--generating transformation, give some examples of its
application, and discuss its general properties.

Stationary Einstein--Maxwell fields without matter sources may be 
parametrized by the metric and electromagnetic fields
\begin{eqnarray}\label{stat}
ds^2 & = & f\,(dt - \omega_i dx^i)^2 - f^{-1}\,h_{ij}\,dx^i dx^j\,,
\nonumber\\ 
F_{i0} & = & \partial_i v\,, \qquad F^{ij} = f\,h^{-1/2}\epsilon^{ijk}
\partial_k u\,,
\end{eqnarray}
where the various fields depend only on the space coordinates $x^i$. In
the case of axisymmetric fields, it is often convenient to choose Weyl
coordinates $\rho$, $z$, $\varphi$ such that
\begin{equation}
\omega_i\,dx^i \equiv \omega(\rho,z)\,d\varphi\,,\qquad h_{ij}\,dx^î\,dx^j
= {\rm e}^{2k(\rho,z)}(d\rho^2 + dz^2) + \rho^2\,d\varphi^2\,.
\end{equation}
The stationary Einstein--Maxwell equations may be reduced to the
three--dimensional Ernst equations \cite{er}
\begin{eqnarray}\label{ernst2}
f\nabla^2{\cal E} & = & \nabla{\cal E} \cdot (\nabla{\cal E} + 
2\overline{\psi}\nabla\psi)\,,\nonumber \\
f\nabla^2\psi & = & \nabla\psi \cdot (\nabla{\cal E} + 
2\overline{\psi}\nabla\psi)\,, \\
f^2R_{ij}(h) & = & {\rm Re} 
\left[ \frac{1}{2}{\cal E},_{(i}\overline{{\cal E}},_{j)} 
+ 2\psi{\cal E},_{(i}\overline{\psi},_{j)}
-2{\cal E}\psi,_{(i}\overline{\psi},_{j)} \right]\,, \nonumber
\end{eqnarray}
where the scalar products and covariant Laplacian are computed with the
reduced spatial metric $h_{ij}$,
the complex Ernst potentials are defined by
\begin{equation}
{\cal E} = f + i \chi - \overline{\psi}\psi\,, \qquad \psi = v + iu\,.
\end{equation}
and $\chi$ is the twist potential 
\begin{equation}\label{twist}
\partial_i\chi = -f^2\,h^{-1/2}h_{ij}\,\epsilon^{jkl}\partial_k\omega_l 
+ 2(u\partial_i v - v\partial_i u)\,.
\end{equation}
These equations are invariant under an SU(2,1) group of transformations
\cite{nk}.

The direct rotation--generating transformation is the product
\begin{equation}\label{sig}
\Sigma = \Pi^{-1}\,{\cal R}\,\Pi
\end{equation}
of three successive transformations, two ``vertical'' transformations
$\Pi\,,\Pi^{-1} \in$ SU(2,1) acting on the potential space, and a
``horizontal'' global coordinate transformation ${\cal R}$ acting
on the Killing 2--plane. The transformation $\Pi$ is the SU(2,1)
involution $({\cal E}, \psi, h_{ij}) \leftrightarrow (\hat{\cal E},
\hat{\psi}, \hat{h}_{ij})$ with
\begin{equation}\label{inv}
\hat{\cal E} = \frac{-1 + {\cal E} + 2 \psi}{1 - {\cal E} + 2 \psi}\,, \quad 
\hat{\psi} = \frac{1 + {\cal E}}{1 - {\cal E} + 2 \psi}\,, \quad
\hat{h}_{ij} = h_{ij}\,.
\end{equation}
Consider the Schwarzschild solution, written in prolate spheroidal coordinates
\cite{zip}
\begin{equation}\label{S}
ds^2 = fdt^2 - f^{-1}m^2\,[dx^2 + (x^2-1)(d\theta^2 + 
\sin^2\theta\,d\varphi^2)]\,,\qquad f = (x-1)/(x+1) 
\end{equation}
(the coordinate $x$ is related to the ``standard'' radial coordinate $r$
by  $x = (r-m)/m$), with the Ernst potentials ${\cal E} = (x-1)/(x+1)$, $\psi =
0$. The action of $\Pi$ leads, after a trivial rescaling of the time
coordinate, to the non asymptotically flat Bertotti--Robinson (BR)
solution \cite{br}
\begin{equation}\label{BR}
d\hat{s}^2 = m^2\, \left[ \,x^2-1)\,d\tau^2 - \frac{dx^2}{x^2-1}  - 
\frac{dy^2}{1-y^2} - (1-y^2)\,d\varphi^2\, \right] \,,
\end{equation}
with the transformed Ernst potentials $\hat{{\cal E}} = -1$, $\hat{\psi} = x$.
More generally, if the initial Ernst potentials are asymptotically
monopole, the transformation $\Pi$ leads to asymptotically BR--like
potentials.

The global coordinate transformation ${\cal R}(\Omega,\gamma)$ is the
product of the transformation to a uniformly rotating frame and of a time
dilation,
\begin{equation}\label{R}
d\varphi = d\varphi' + \Omega\gamma\,dt'\,, \qquad dt = \gamma\,dt'\,.
\end{equation}
In the case of electrostatic solutions with $\hat{{\cal E}}$ and $\hat{\psi}$
real, ($\hat{\omega} = 0$), the frame rotation gives rise to an induced
gravimagnetic field $\hat{\omega}'$ as well as to an induced magnetic
field. However this transformation does not modify the leading asymptotic
behavior of the BR metric or of asymptotically BR--like metrics. Because
of this last property, the last transformation $\Pi^{-1}$ in (\ref{sig})
then leads to asymptotically flat, but complex, Ernst potentials
corresponding to a monopole--dipole solution.

As shown in \cite{kerr} the transformation $\Sigma$ leads, for the special
choice $\gamma = (1+m^2\Omega^2)^{-1/2}$, from the Schwarzschild solution 
to the Kerr solution. For other values of $\gamma$ the Kerr--Newman family
of solutions is obtained. 
Another example is that of the Voorhees--Zipoy family of static vacuum
solutions \cite{zip}, depending on a real parameter $\delta$. The action
of the transformation $\Sigma$ leads to new rotating solutions \cite{kerr}
with dipole magnetic moment and quadrupole electric moment which are
different from the discrete ($\delta$ integer) Tomimatsu--Sato \cite{ts}
family of vacuum rotating solutions. Yet another example is the generation
of spinning ring solutions of the Einstein--Maxwell equations from static
ring wormhole solutions \cite{lring}.

In the case of a generic axisymmetric electrostatic solution (${\cal E}$,
$\psi$ real), the transformed BR--like Ernst potentials $\hat{{\cal E}}$,
$\hat{\psi}$ satisfy the real Ernst equations, 
\begin{equation}
\nabla(\rho\hat{f}^{-1}\nabla\hat{{\cal E}}) = 0\,,\qquad 
\nabla(\rho\hat{f}^{-1}\nabla\hat{\psi}) = 0\,,
\end{equation}
which imply the existence of dual Ernst potentials $\hat{\cal F}$,
$\hat{\phi}$ such that 
\begin{equation}\label{dual}
{\hat{\cal F}}_{,m}  = \rho\hat{f}^{-1}\,\epsilon_{mn}{\hat{{\cal E}}}_{,n}\,,
\quad {\hat{\phi}}_{,m} = \rho\hat{f}^{-1}\,\epsilon_{mn}{\hat{\psi}}_{,n}
\end{equation}
($m$,$n$ = 1,2, with $x^1=\rho$, $x^2=z$). It may then be shown that the 
transformation ${\cal R}$ with $\gamma = 1$ transforms the potentials 
($\hat{{\cal E}}$, $\hat{\psi}$, ${\rm e}^{\hat{2k}}$) into
\begin{eqnarray}\label{crank}
\hat{{\cal E}}' & = & \hat{{\cal E}} + 2i\Omega\,(z + \hat{\cal F} + 
\hat{\psi}\hat{\phi}) - \Omega^2 (\rho^2/\hat{f} +
\hat{\phi}^2)\,,
\nonumber \\ 
\hat{\psi}' & = & \hat{\psi} + i\Omega\hat{\phi}\,, \qquad {\rm e}^{2\hat{k}'}
= (1 - \Omega^2\rho^2/\hat{f}^2)\,{\rm e}^{2\hat{k}}\,.
\end{eqnarray}
From these one may write down the asymptotically flat potentials
${\cal E}'$, $\psi'$, from which the rotating metric $g_{\mu\nu}'$ and the
rotating electromagnetic potentials $A_{\mu}'$ may be obtained by solving duality
equations. I give only here the expressions of the functions entering the
Weyl form of this metric (again for $\gamma = 1$):
\begin{eqnarray}\label{spin}
f' & = & (|\hat{F}|^2/|\hat{F}'|^2)\lambda\,f\,, \qquad {\rm e}^{2k'} =   
\lambda\,{\rm e}^{2k}\,, \nonumber \\
\partial_m\omega'& = & \Omega^{-1}|\hat{F}'|^2 \partial_m\lambda^{-1} 
- (2\rho/\hat{f})\lambda^{-1}\epsilon_{mn}
{\rm Im}(\hat{\overline{F}}'\partial_n\hat{F}')\,,
\end{eqnarray}
with $|\hat{F}'| \equiv (1/2)|1 - \hat{{\cal E}}'+ 2\hat{\psi}'| = 1/|F'|$,
$\lambda \equiv (1-\Omega^2\rho^2/\hat{f}^2)$. 

What are the general properties of the rotating Einstein--Maxwell fields
thus generated? It can easily be checked that they are regular on the axis
$\rho = 0$,
\begin{equation}
{\rm e}^{2k'} = 0\,, \qquad \partial_z\omega' = 0\,,
\end{equation}
if the original static fields are regular. Another obvious property is
the existence of stationary limit surfaces $f'(\rho,z) = 0$ for
$\hat{f}(\rho,z) = \pm\Omega\rho$. Near such a zero of $f'$, the rotating
metric 
\begin{equation}
ds'^2 \simeq \mp 2\rho\,dt\,d\varphi 
\mp(\hat{F}'^2/\Omega\rho)({\rm e}^{2k}(d\rho^2 + dz^2) +
\rho^2\,d\varphi^2) 
\end{equation}
is non--degenerate.

It also seems that the horizons of the static solution (zeroes of $f$)
generically carry over to the resulting rotating solutions, although a
fully convincing proof is not available at present (Weyl coordinates are
not well adapted to this purpose). An illustration of this horizon
conservation is the fact that the transformation $\Sigma$ (with arbitrary
$\Omega$ and $\gamma$) transforms an extreme Reissner--Nordstr\"{o}m black
hole ($M^2 = Q^2$) into an extreme Kerr--Newman black hole ($M'^2 = Q'^2 +
a'^2$).

The rotating metric (\ref{spin}) may present Kerr--like ring singularities
corresponding to the zeroes of the function $|\hat{F}'|^2(\rho,z)$. In the
plane--symmetric case, these rings are located in the plane $z = 0$ 
(${\rm Im}\hat{F}'= 0$) with radii given by the solutions of the equation
\begin{equation}\label{sing}
2\,{\rm Re}\hat{F}' = (1+\hat{\psi})^2 - \hat{f}' =  0 \qquad {\rm for} z=0
\end{equation}
($\hat{f}' \equiv \hat{f} - \Omega^2\rho^2/\hat{f}$). In the case of the
solutions studied in \cite{lring}, these unwanted ring singularities may
be avoided by suitably choosing the parameters of the static solution
and/or the parameter $\Omega$.

The transformation $\Sigma$ does not generically commute with SU(2,1)
transformations. This raises the question of classifying inequivalent
solutions generated from a given static solution by transformations ${\cal
U}\Sigma{\cal U}^{-1}$, with ${\cal U} \in {\rm SU(2,1)}$. A related
question is that of the precise connection of the direct spin--generating
method presented here with other spin--generating techniques.

\end{document}